\documentclass[11pt]{article}
\usepackage[letterpaper, left=2cm, right=2cm, top = 1truein, bottom = 1truein]{geometry}

\usepackage{amsmath}
\usepackage{amssymb}
\usepackage{amsthm}
\usepackage{tgbonum}
\usepackage{epsfig}
\usepackage{epstopdf}

\usepackage{cite}
\usepackage{tikz}
\usepackage{color}
\usepackage{pgfplots}
\usepackage{graphicx}
\usepackage{color}
\usepackage{bbm}
\usepackage{bm}
\usepackage{showlabels}
\showlabels{label, cite, thm}

\DeclareMathAlphabet{\mathpzc}{OT1}{pzc}{m}{it}

\newcommand{\zero}{{\bm{0}}}

\newtheorem{thm}{Theorem}
\newtheorem{lem}[thm]{Lemma}
\newtheorem{prp}[thm]{Proposition}
\newtheorem{cor}[thm]{Corollary}
\newtheorem{dfn}[thm]{Definition}

\newtheorem{obs}[thm]{Observation}

\newcommand{\Real}{\mathbb{R}}

\def\be{\begin{equation}}
\def\ee{\end{equation}}

\def\Int{\mathbb{Z}}
\def\Nat{\mathbb{N}}
\def\Comp{\mathbb{C}}
\def\Real{\mathbb{R}}

\def\bx{\mathbf{x}}

\def\tr{\mathtt{tr}}

\def\basis{\mathcal{B}}
\def\hilb{\he_L}

\def\coin{\mathbf{U}}
\def\bk{\mathbf{k}}
\def\bl{\mathbf{l}}
\def\bz{\mathbf{z}}

\def\bj{\mathbf{j}}
\def\qstep{\mathbf{S}}
\def\b0{\mathbf{0}}
\def\torus{\mathbb{T}}
\def\torL{\torus_L}
\def\ch{\mathbf{P}}

\def\spectral{\Sigma}

\def\gauss{\bm{\mathtt{G}}}
\def\limmes{\bm{\gamma}}
\def\simplex{\bm{\Delta}}

\def\jt{\mathbf{j}}

\def\ex{\mathbb{E}}

\def\hil{{{H}}}
\def\basis{\bm{C}}
\def\hilb{\hil_L}
\def\coin{\bm{U}}
\def\bk{{\bm{k}}}
\def\bl{\bm{l}}

\def\bj{\bm{j}}
\def\vbx{\overline{\bx}}
\def\vxi{\bm{\xi}}
\def\veta{\bm{\eta}}

\def\qstep{\bm{S}}
\def\b0{\mathbf{0}}
\def\torus{\mathbb{T}}
\def\torL{\torus_L}
\def\etorus{\bar{\torus}}
\def\ch{\mathbf{P}}

\def\spectral{\mathbf{S}}
\def\smspectral{{\spectral^\circ}}
\def\sing{\Sigma}
\def\smstorL{\torL^\circ}

\def\gauss{\bm{\mathtt{G}}}
\def\limmes{\bm{\gamma}}
\def\simplex{\bm{\Delta}}

\def\jt{\mathbf{j}}

\def\dia{{\Delta}}
\def\eye{\mathbf{I}}
\def\su{\mathrm{SU}}

\usepackage{etoolbox} \makeatletter
\patchcmd{\@maketitle}{\begin{center}}{\begin{flushleft}}{}{}
\patchcmd{\@maketitle}{\begin{tabular}[t]{c}}{\begin{tabular}[t]{@{}l}}{}{}
\patchcmd{\@maketitle}{\end{center}}{\end{flushleft}}{}{}

\title{On Random Quantum Random Walks}

\date{\today}
\author{Yuliy Baryshnikov\thanks{Supported in part by NSF DMS grant 1622370} \thanks{\tt publish.illinois.edu/ymb}\\\\
  \parbox{.6\textwidth}{\small
    Department of Mathematics and Electrical and Computing Engineering,
  University of Illinois, Urbana, IL, USA, and\\
  IMI, Kyushu University, Japan.}
  }

\begin{document}
\maketitle
\begin{abstract}
Quantum random walks, - coined, lattice ones, in our case, - have ballistic behavior with fascinating asymptotic patterns of the amplitudes. We show, that averaging over the coins (using the Haar measure), these patterns blend into a simple pattern. Also, we discuss the localizations of such quantum random walks, and establish some strong constraints on the achievable speeds. 
\end{abstract}
\section{Introduction}
Quantum random walks (QRWs) is a class of unitary evolution operators
that combine the geometry of the underlying space, indexing the
positions of the walk, and the internal state dynamics. The literature
on quantum random walks is vast, and still is growing, - for more
recent survey see, e.g. \cite{venegas-andraca_quantum_2012}.

We deal here with the so-called {\em coined quantum random walks} in
discrete time on a lattice. The coin remains the same, for all time
and space positions, and this translation invariance of the walks we
consider allows one to resort to one or another version of Fourier
transform, and to understand the corresponding evolution quite
precisely.

In particular, one knows that the behavior of such QRWs is {\em
  ballistic}, that is the amplitudes are spreading on a linear (in time) scale over the lattice.
These amplitudes exhibit some rather intricate dependence on the data defining the walks,
addressed in quite a few papers.

One of the features one can observe by numeric experiments for the 1- and
2-dimensional lattices, is that the amplitudes, and the corresponding
probabilities, form fascinating moire-like patterns, depending on the
coin and the jump map (see the definitions below). One of the
motivation for this note was to understand the {\em typical}
behavior of such patterns.

More specifically, the question we sought
to answer is: when the amplitudes are averaged over the random coin
and the initial internal state, what are the resulting probabilities to
find the the particle in a given position?

This randomness is different from often considered random coin model,
where the realization of the coin differ from time to time, or from
site to site.  (In this case, the behavior is not ballistic, but
rather diffusive, see e.g. \cite{ahlbrecht_asymptotic_2011}.) In our
situation, for a given coin we observe some ballistic propagation
pattern, and then average these patterns over the coins.

The natural probability measure on the unitary coins is the Haar
measure over the unitary group $\su(c)$.  It turns out that in this
case the question can be answered precisely (see Theorem
\ref{thm:main}): the averaged probability is the push-forward under
the {\em jump map} of the uniform measure on the simplex spanned by
the internal states. The intricate patterns boringly add up to a
spline.

Besides that result, a few other novel (I believe) results are
presented in this note. Thus, in the section \ref{sec:prob} we sketch
a new proof of the characterization of the weak limits of the (scaled)
position of the QRW in terms of the Gauss map. Further, in section
\ref{sec:local} we address the localization of QRWs, showing that the
strong localization is equivalent to the weak one, and prove that the
localization speeds are quite constrained.

\subsection{Translation Invariant Lattice QRWs}
To define a (translation invariant) quantum random walk on a lattice one needs the following data:
\begin{itemize}
\item the lattice $L\subset\Real^d$ of rank $d$ (in what follow, we will be just assuming $L=\Int^d$, although sometimes it is convenient to use a lattice possessing a different symmetry group);
\item  the {\em chirality space}, that is a $c$-dimensional Hilbert space
    $\hil\cong\Comp^c$ with a fixed orthonormal basis
    $\basis:=\{e_1,\ldots,e_c\}$;
\item  the {\em coin}: a unitary operator $\coin$ acting on the chirality space $\hil$;
\item the {\em jump map}: a mapping $\jt:\basis\to L$ (we will assume, without loss of generality, that the jumps
    $j_k=\jt(e_k), k=1,\ldots,c$ span $\Real^d$ affinely; otherwise one can just restrict to the sublattice spanned by the jumps, and a smaller ambient space.
\end{itemize}

Tensoring $l_2(L)$ with $\hil$ results in the Hilbert space $\hilb:=l_2(L)\otimes\hil$ with the basis $|\bk,v\rangle, \bk\in L, v\in\basis $.

Notation: we will be using $\langle \cdot,\cdot\rangle$ for the Hermitean product in both $\hil$ or $\hilb$, whenever this does not lead to a confusion.


\subsubsection{Defining Quantum Random Walk}

The quantum random walk associated with these data is the discrete time unitary
evolution on $\hilb$ resulting from the composition of two operators,
$\qstep=S_2\circ S_1$, which are defined, in turn, as follows:

The operator $S_1$ applies the coin at each site of the lattice, that is 
$$
S_1=\mathtt{id}_{l_2(L)}\otimes U,
$$

(i.e. $\coin$ acts on each $\hil_{\bk}, \bk\in L$ independently).

The second operator is the composition of shifting each of the ``layers''
$l_2(L)\otimes e_k, k=1,\ldots,c$ by $j(e_k)$
\[
S_2:\bk\otimes e_k \mapsto (\bk+\jt(e_k))\otimes e_k.
\]

It is immediate that both $S_1, S_2$ are unitary, as is their composition.

\subsubsection{Evolution of Coined QRWs}

The evolution defined by $\{\qstep^T\}_{T\in\Nat}$ exhibits {\em
  ballistic behavior}: the support of the amplitudes
in $L$ grows linearly with $T$ (see examples below).

From the construction it should be clear that the matrix elements
$$
a_T(\bl, u;\bk, v):=\langle \bk\otimes v|\qstep^T \bl\otimes u\rangle
$$

depend on $\bl$ and $\bk$ only through $\bl-\bk$, and vanish if
$\bl-\bk$ cannot be represented as a sum of $T$ lattice vectors from
the jump set $j(\basis)$. In particular, the matrix elements vanish
when $\bl-\bk$ is outside of the scaled by $T$ convex hull of the jump
vectors $\jt(\basis)$ which we denote as
$$
\ch=\ch_{\jt}:=\mathtt{conv}\left\{\jt(e_k), k=1,\ldots,c\right\}.
$$

We will use the shorthand
\be
A_T^{u,v}(\bk):=a_T(\zero, u;\bk, v):=\langle\zero\otimes v,\qstep^T \bk\otimes u\rangle
\ee

for the amplitudes of the quantum random random walk starting at site
$\zero$ in the internal state $u$. Further, we denote by $A_T(\bk)$ the
corresponding operator $\hil_\zero\to\hil_{\bk}$, whose matrix
coefficients are $A_T^{u,v}(\bk)$.

We will be mostly interested in the {\em probabilities} (of
transitions between states)
$$
p_T^{u,v}(\bk):=|A_T^{u,v}(\bk)|^2.
$$

The unitarity of $\qstep$ implies that $\{p_T^u(\bk,v)\}_{\bk\in L,v\in\basis}$ is a probability
distribution on the basis $\{|\bk, v\rangle\}_{\bk\in L, v\in \basis}$
of $\hilb$ for any $T\geq 0$ and norm one $u\in\hil$.

By construction, it is also clear that the amplitudes $a_T(\zero,
u;\bk, v)$ belong to the (dense) subspace of $\hilb$ of vectors with
all almost all components zero.

\subsection{Examples}\label{subsec:examples}

In this section we will look at a few examples of QRWs in $d=2$.
\subsubsection{Hadamard Coin}
A popular class of examples uses the {\em Hadamard (or Grover) coins} given by the real matrices
\[U=\eye-({2}/{c})\mathbf{O},
\]
where $\eye$ is the identity matrix, and $\mathbf{O}$ is the $c\times c$ matrix with all components equal to $1$.

For $c=4$, it is 
given by
\be\label{eq:hadamard}
U={1\over 2} \begin{pmatrix} 1 & -1 & -1 & -1 \\ -1 &
1 & -1 & -1 \\ -1 & -1 & 1 & -1 \\ -1 & -1 & -1 & 1 \\
\end{pmatrix}
\ee

In the standard setting, the jumps are the steps to the neighboring
sites on the $2$-dimensional integer grid, so that the jump map
takes the basis vectors into $\jt(\basis)=\{(0,\pm 1),(\pm 1,0)\}$.

In this case, the amplitudes have asymptotic support localized in the
circle inscribed into the diamond $\ch$, and has been thoroughly
analized, see e.g. \cite{baryshnikov_two-dimensional_2011}.

\begin{figure}[h!]

 \begin{center}
\includegraphics[height=1.5in]{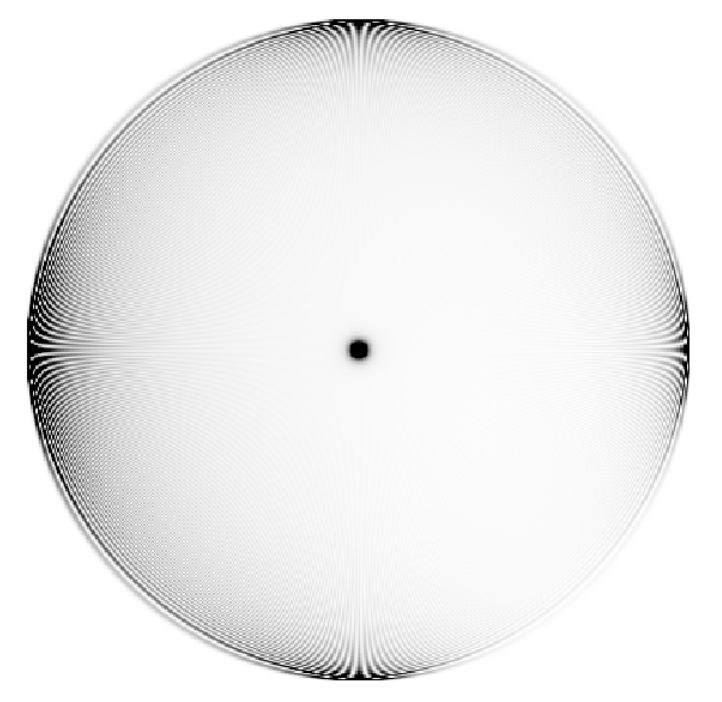}\hfill
\includegraphics[height=1.5in]{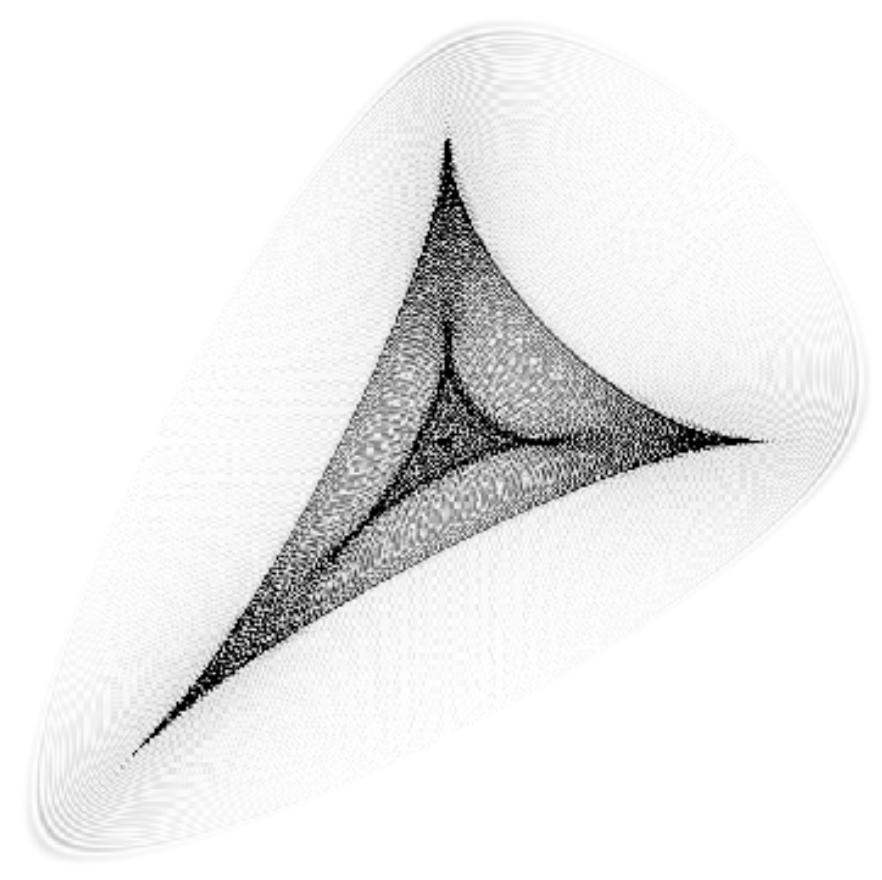}\hfill
\includegraphics[height=1.5in]{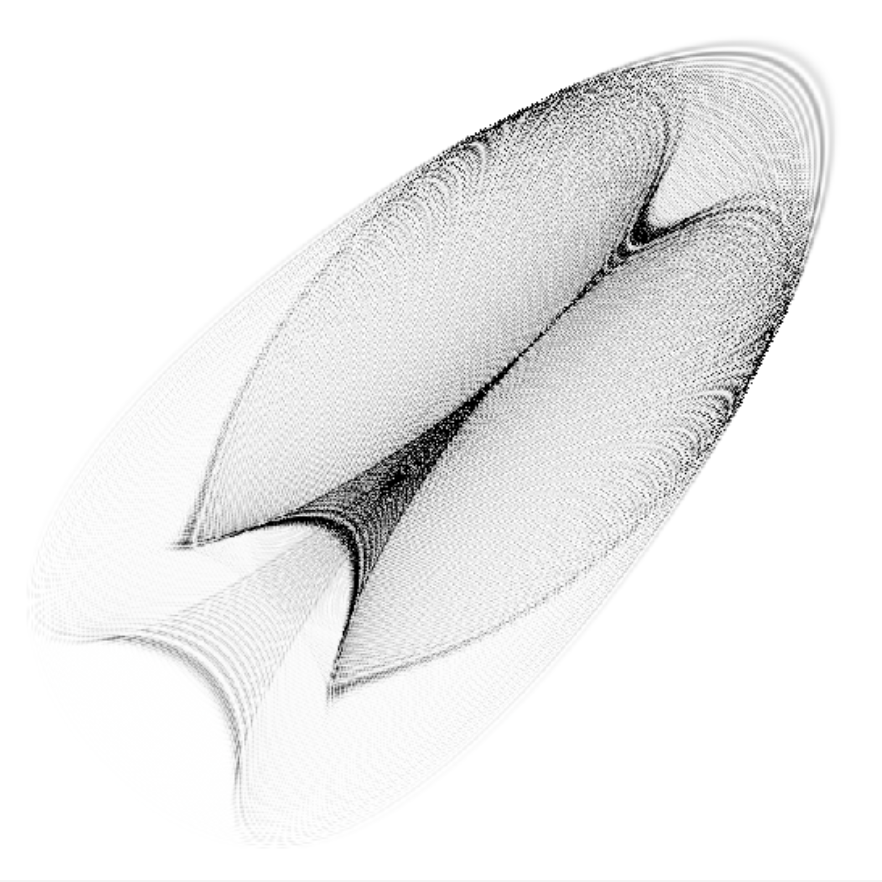}\hfill
\includegraphics[height=1.5in]{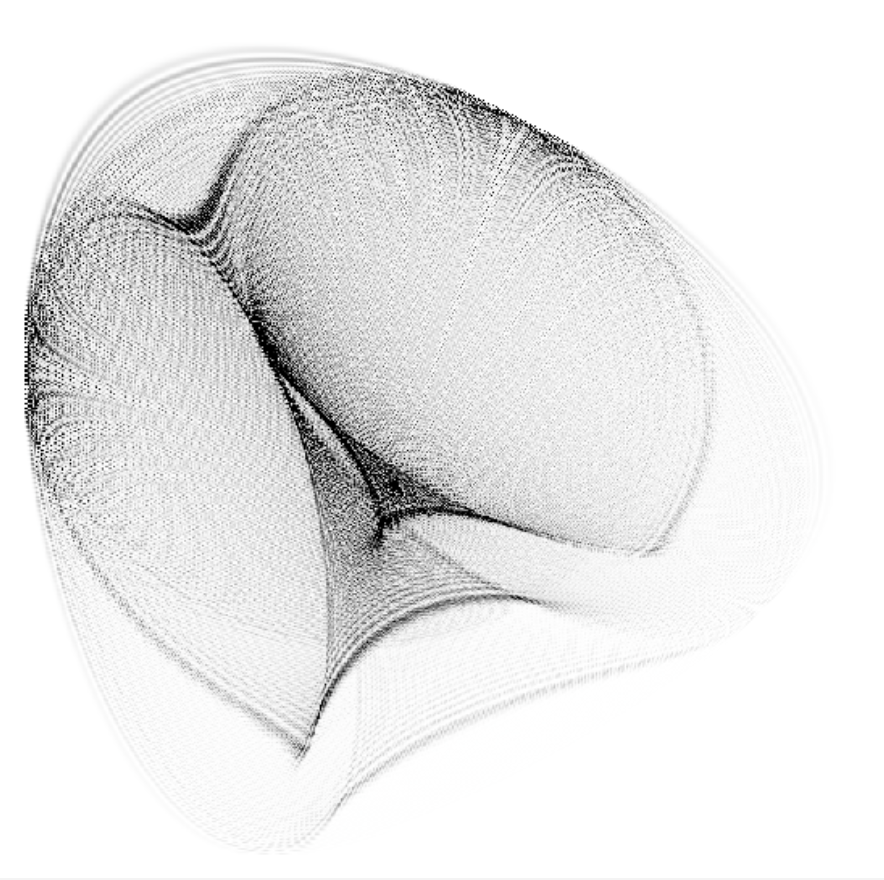}
\caption{Probability distributions for QRWs with $d=4$ Hadamard coin and jump vectors $\jt_1,\ldots,\jt_4$, as in section \ref{jump}.  
}
\label{fig:hadamard}
\end{center}
\end{figure}

Switching to the jump map sending the basis to
$\{(0,0),(0,1),(1,0),(1,1)\}$ leads to an essentially equivalent
picture: the difference is that the support of the amplitudes now is
not the (shifted) even sublattice of $\Int^2$, but the entire integer
lattice, and the footprint acquires drift: it is centered at
$(1/2,1/2)$. The simulated amplitudes (or rather the corresponding
probabilities averaged over all possible initial internal states) are shown
on the left display of Figure \ref{fig:hadamard} - after 400 steps of
the walk.

One can, of course, use some completely different jump maps for the same
coin. The remaining three displays of Figure \ref{fig:hadamard} show the probabilities of the QRW using the coin \eqref{eq:hadamard} for the
jump map sending the basis of $\hil$ to other collections of vectors (spanning
$L$). The jump maps for the four displays (left to right) are given by
\begin{itemize}\label{jump}
\item $\jt_1(\basis)=\{(0,0),(0,1),(1,0),(1,1)\}$;
\item $\jt_2(\basis)=\{(0,0),(2,1),(1,2),(1,1)\}$;
\item $\jt_3(\basis)=\{(0,0),(2,1),(1,2),(2,2)\}$ and
\item $\jt_4(\basis)=\{(0,0),(2,0),(2,1),(1,0)\}$.
\end{itemize}

Worth remarking that in the second from left display, one can apply an affine transformation taking the standard square lattice (used to render the picture) to the standard hexagonal one. If under this tranformation the jump vectors are sent to the zero vector and the three shortest vectors of the hexagonal lattice, then the probabilities would become symmetric with respect to all possible Euclidean automorphisms of the lattice, for obvious reasons (as the coin is invariant with respect to swapping the elements of the basis).

\subsubsection{Random Unitary Coin}

If one chooses a generic coin, the amplitudes change dramatically.
Below are the results of the simulations for the chirality space of the same dimension $c=4$, but for a
(randomly generated) unitary matrix $U$: 
\be\label{eq:ucoin}
{\small
U=\left(\begin{array}{rrrr}
{\mathtt-0.331759+0.069082 i}&  0.471768+0.231231 i &  -0.278617-0.583254 i & -0.425926+0.099521 i\\
{\mathtt  0.368644-0.479381 i}&  -0.113567+0.513171 i&   -0.443628+0.254345 i &  -0.218580+0.220861 i \\
{\mathtt  0.169821-0.199957 i}&   0.206809-0.447012 i&    0.177488+0.271570 i  &  -0.723490-0.244741 i\\ 
{\mathtt -0.654156+0.150721 i}&  -0.444209+0.088396 i&  -0.315095+0.340811 i & -0.234176-0.271940 i\\
\end{array}\right)
}
\ee
As before, we take $T=400$ time
steps, and the jump maps $\jt_1,\ldots,\jt_4$ are matching those for the Hadamard example.

\begin{figure}[h!]
  \begin{center}
\includegraphics[height=1.5in]{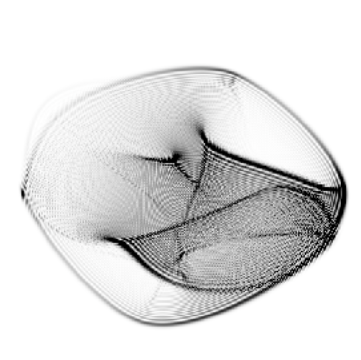}\hfill
\includegraphics[height=1.5in]{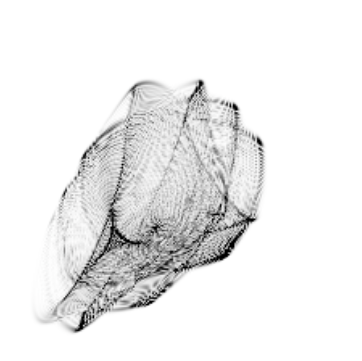}\hfill
\includegraphics[height=1.5in]{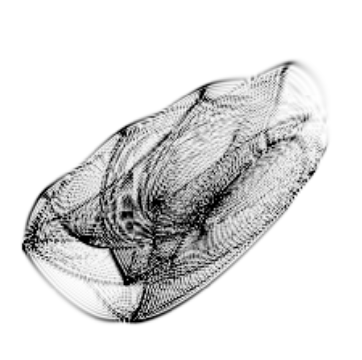}\hfill
\includegraphics[height=1.5in]{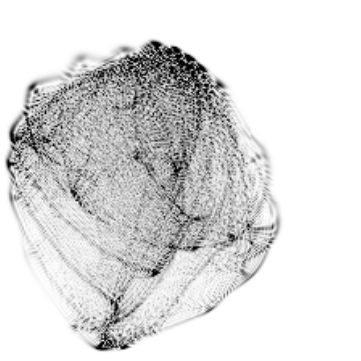}
\caption{Probability distributions for QRWs with $d=4$, unitary coin \eqref{eq:ucoin} and jump maps $\jt_1,\ldots,\jt_4$, as in section \ref{jump}.  
}
\label{fig:hadamardrand}
\end{center}
\end{figure}

Again, we see that the amplitudes depend strongly on both the coin and the jump map.

\section{Amplitudes as Oscillating Integrals}

As the numeric simulations show, the coefficients $A_{T}(\cdot)$
exhibit intricate interference patterns, supported within the
polytope $T\ch$, scaled by $T$ convex hull of the jump vectors. Experiments show that the support of the probability distribution  $p_T$ is a proper subset of $T\ch$. However, as we will see, the amplitudes do not vanish at the lattice points of $T\ch$ which are outside of the visible support of $p_T$, but rather are exponentially (in $T$) small there.

Let $\vbx=(\bx_1,\ldots,\bx_d)$ be the vector whose components we
interpret (for now) as symbolic variables. We associate to the
amplitudes $A_T^{u,v}(\cdot)$ their generating (in general, Laurent) polynomials
\[
A_T^{u,v}(\vbx):=\sum_{\bk\in L} A_T^{u,v}(\bk) \vbx^\bk,
\]
where we use the shorthand 
\[
\vbx^\bk=\prod_{l=1}^d x_l^{k_l}.
\]

We can regard $A_T^{\cdot,\cdot}(\vbx)$ as a matrix with coefficients in
$\Comp((\vbx))$, the ring of Laurent polynomials in variables
$\bx_1,\ldots, \bx_d$.

The following Proposition is a straightforward corollary of the
definitions:
\begin{prp}
For any $T\geq 0$,
\[
A_T(\vbx)=M(\vbx)^T,
\]
where $M(\vbx)=\dia(\vbx)\coin$, and $\dia(\vbx)$ is the diagonal matrix with $\dia(\vbx)_{l,l}=\vbx^{\jt(e_l)}$.
\end{prp}

\subsection{Complex Variables}
From now on, we interpret all $\bx_l,\bz$ as complex variables.

In this case the matrix $M(\vbx)$ becomes a matrix with complex coefficients.

Let $L^*$ be the lattice of linear functionals on $\Real^d$ taking integer values on $L$, and $\torus_L\cong\Real^d/L^*$ be the $d$-dimensional torus (of
characters on $L$). We can identify $\torus_L$ with the collection of
vectors $\vbx\in\Comp^d: |\bx_l|=1, l=1,\ldots,d$. The product of
$\torus_L$ with the unit circle (coordinatized by $\bz$) is called the {\em extended torus}:
\[
\etorus=\torus_L\times \torus^1=\left\{(\vbx,\bz): \vbx\in\torus_L, |\bz|=1\right\}.
\]

As $\dia(\vbx)$ is unitary for $\vbx\in\torL$, so is the matrix $M(\vbx)$.

The precise structure of matric coefficients of $A_T$ is best understood in terms of the {\em spectral surface}.

\begin{dfn}
Consider $M(\vbx)$ as the function on $\torL$ with values in the unitary operators. Then 
\[
\spectral:=\{(\vbx,\bz): \bz \mbox{ is an eigenvalue of } M(\vbx)\}
\]
is called {\em the spectral surface}. 
\end{dfn}

The following is immediate:
\begin{lem}
The spectral surface is (real) algebraic,
and its projection $p:\spectral\to\torL$ is a $c$-fold branching covering (counted with multiplicities).
\end{lem}
\begin{proof}
Indeed, the spectral surface is given by the zero set of the Laurent
polynomial
\[
\det(z\eye-M(\vbx)),
\]
and the spectrum of the unitary matrix $M(\vbx)$ belongs to the unit circle.
\end{proof}

We will denote the fiber of the projection to $\torL$ as 
\[
\spectral(\vbx):=p^{-1}(\vbx)\cap \spectral=\{\bz: (\vbx,\bz)\in\spectral\}.
\]

The spectral theorem implies that one can associate to any point $(\vbx,\bz)$ on the spectral surface the projector $P((\vbx,\bz)$ to the corresponding eigenspace, so that
\[
M(\vbx)=\sum_{\bz\in \spectral(\vbx)} \bz P(\vbx,\bz).
\]
These projectors $P((\vbx,\bz)$ are orthogonal for different $\bz\in \spectral(\vbx)$, and sum up to the unity,
\[
\sum_{\bz\in \spectral(\vbx)} P(\vbx,\bz)=\eye.
\]
Therefore,
\be\label{eq:amplitudes}
A_T(\vbx)=\sum_{\bz\in \spectral(\vbx)} \bz^T P(\vbx,\bz).
\ee

At the points where the spectral surface is smooth, $\bz$ can be
represented (locally) as a function of $\vbx$. If we define
$\sing=\sing(\coin)\subset\spectral$ as the set of singular points of
$\spectral$, one has
\begin{prp}
  The set $S$ is a real algebraic subvariety of $\spectral$, such that
  $\smspectral=\spectral-\sing$ is everywhere dense in $\spectral$.
  The image of the projection of $\sing$ to $\torL$ is
  nowhere dense.
\end{prp}

Remark that the smoothness of the spectral surface at $(\vbx,\bz)$
does not imply necessarily that the rank of the projector
$P(\vbx,\bz)$ at the point is $1$: one can have a smooth component in
$\spectral$ of multiplicity $>1$. However, for a generic $\coin$ this
does not happen.

\subsection{Amplitudes as Oscillating Integrals}

To recover the matrix elements from the expression \eqref{eq:amplitudes} we apply the Cauchy formula (or, equivalently, invert the Fourier transform):
\begin{prp}
  The amplitudes $A_T(\bk)$ are given by
\be
A_T(\bk)=\frac{1}{(2\pi i)^d}\int_{\torL} \vbx^{-\bk} M(\vbx)\frac{d\vbx}{\vbx}=\frac{1}{(2\pi i)^d}\int_{\torL}  \sum_{\bz\in\spectral(\vbx)} \bz^T\vbx^{-\bk} P(\vbx,\bz)\frac{d\vbx}{\vbx}.
\ee
\end{prp}

Introducing the logarithmic coordinates on the extended torus $|\bx_l|=1,l=1,\ldots, d, |\bz|=1$,
\[
\bx_k=\exp(i\xi_k), \bz=\exp(i\zeta),
\]

we obtain
\be
\label{eq:osc1} A_T(\bk)=\int_{0\leq\xi_k\leq 2\pi }
\sum_{\zeta\in\spectral(\vxi)} e^{iT(\zeta-(\rho,\vxi))} P(\vxi,\zeta)
d\vxi,
\ee

where we denote by $\rho=\bk/T$ the rescaled sites in $L$, and by
$\vxi=(\xi_1,\ldots,\xi_d)$. (We will retain the notation
$\spectral(\vxi), P(\vxi,\zeta)$ for the spectral surface etc in the
logarithmic coordinates, whenever this does not lead to confusion.)

The identity \eqref{eq:osc1} expresses the amplitudes as {\em oscillating integrals}.  Namely, denote by
$\sing_{\vxi}=p(\sing)$ the projection of the singular set of
$\spectral$ to the $\torL$ (which is a nowhere dense, closed
semialgebraic subset of $\torL$), and let $\smstorL:=\torL-\sing_{\vxi}$ be its complement: an open, everywhere dencse subset of $\torL$, such that the fiber $p^{-1}(\vxi)$ for any point $\vxi\in\smstorL$ intersects the spectral surface only at the smooth points.

In a vicinity $U$ of such a point
$\vxi_*$, the branches of $\zeta$ can be
represented locally as functions of $\vxi$, so that the corresponding
contribution to the integral \eqref{eq:osc1} becomes
\be\label{eq:osc2} \sum_m
\sum_{\zeta_\alpha\in\spectral(\vxi_*)}\int_U
e^{iT(\zeta_\alpha(\vxi)-(\rho,\vxi))} P_\alpha(\vxi) s_m d\vxi, \ee
where we denote by $P_\alpha(\vxi)=P(\vxi,\zeta_\alpha)$; the
external summation is over the open vicinities of an open covering $\bigcup_m
U_m=\smstorL$, and $s_m$ is a subordinated partition of the
unity.

This representation allows one to use various tools from the
theory of oscillating integrals to explore large $T$ asymptotic behavior of the amplitudes.

Thus a standard result \cite{AVG2} implies that if for some $\rho$ the phase 
\[
\zeta_\alpha(\vxi)-(\rho,\vxi)
\]
has no critical points on any of the branches $\zeta_\alpha$, the integral
decays faster than any power of $T$.

Therefore, if $\rho$ {\em is  not} in the range
of $d\zeta_\alpha$, for any $\alpha$, then the amplitudes are decaying
superpolynomially (in fact, exponentially fast) at the indices
$\bk\approx\rho T$, as $T\to\infty$.

If the phase $\zeta_\alpha-(\rho,\vxi)$ does have a critical point, it
is, for a generic $\rho$, a Morse one, i.e. has non-degenerate
quadratic part. In this case, again according to the standard results \cite{AVG2},
the amplitudes decay as $T^{-d/2}$ (this meshes well with the fact
that squared amplitudes behave generically as $T^{-d}$, as they
represent a discrete probability distribution supported by a subset of
the lattice of cardinality $\Theta(T^d)$).

Looking deeper, near a typical point of the boundary of the essential support
of the amplitudes, they are given by an Airy type integral. One can find also the
Pearcey integrals (at isolated points, for the $d=2$-dimensional QRWs), and further oscillating integrals depending on
parameters. We will expound elsewhere on the relations between the properties of the quantum random walks and the complexity of the oscillating integrals appearing in the asymptotic expansions of their amplitudes.

\section{Probability Measures associated with a QRW}\label{sec:prob}

Like the amplitudes, the (discrete) nonnegative measures 
$$
p^{u,v}_T(\bk)=|A_T^{u,v}(\bk)|^2
$$ oscillate wildly for large $T$. However, after rescaling they converge
weakly to a well-defined probability measure. This probability measure
has a nice characterization described first in
\cite{grimmett_weak_2004}.

\subsection{Gauss Map}
For a smooth point $(\vxi,\zeta)\in\smspectral$ of the spectral
surface, we denote by $\gauss(\vxi,\zeta)$ the differential of
$\zeta(\vxi)$, an implicit function parameterizing the
branch of the spectral surface passing through $(\vxi,\zeta)$. Note that all tangent spaces to points of
$\etorus=\torL\times\torus$ can be canonically identified between themselves, and therefore with a fixed Euclidean space
$\Real^d\times\Real$. Hence we can regard $\gauss(\vxi,\zeta)$ as a covector on $\Real^d$.

We will be referring to $\gauss$ as the {\em
  Gauss map}. The Gauss map is defined on a dense subset $\smspectral$ of
$\spectral$.

Fix $u\in \hil$, the initial state.
We will denote by 
\[
p_T^u(\bk):=\int_{|v|=1} p_T^{u,v}(\bk)\sigma(dv)=\sum_{v\in\basis} p_T^{u,v}(\bk)
\]
the total probability measure corresponding to initial state $|\zero,u\rangle, |u|=1$, obtained by the averaging the probability $p^{u,v}_T(\bk)$ over all possible finite states $v$.

We are interested in the weak limits of the rescaled (discrete) probability measures
\[
\pi_T^u(\rho):=p_T^u(T\rho).
\]
These measures describe discrete random variables supported by $\ch\cap L/T$, the
intersection of convex hull of the jump vectors with the rescaled
lattice $L/T$.

\subsection{Weak Limits}
The limiting behavior of these measures is given by the
Theorem \ref{thm:weak}.  A version of this result was proven in \cite{grimmett_weak_2004}) using momenta. HereI will sketch an alternative proof relying, again, on the tools of the theory of oscillating integrals.

\begin{thm}\label{thm:weak}
  Define the nonnegative densities $s^{u}$ on (the dense nonsingular
  part of) the spectral surface $\spectral$ as
  \[
    s^{u}(\vxi,\zeta):=\langle u, P(\vxi,\zeta) u\rangle d\vxi.
  \]
  Then, as $T\to\infty$, $\pi^u_T$ weakly converges to  $\gauss_*s^{u}$, the
  push-forward of the density $s^u$ under the Gauss
  map.
\end{thm}

\begin{proof}[Sketch of the proof]
We will use the trick introduced by Dusitermaat in \cite{duistermaat}.

To prove weak convergence of the probability measures, it is enough to
prove the convergence of the integrals
\be\label{eq:convo}
\int
h(\rho)p_T^u(T\rho) \to \int h(\rho)\gauss_*s^{u}(\rho)
\ee
for smooth compactly supported test functions $h$.

We will the representation \eqref{eq:osc2}. Substituting, we obtain

\begin{align}
   p^{u,v}_T(\bk)=&|A^{u,v}_T(\bk)|^2=\\ =\int\int&
   \sum_{\zeta_\alpha\in\spectral(\vxi),\zeta_{\alpha'}\in\spectral(\vxi')}e^{iT(\zeta_\alpha(\vxi)-\zeta_{\alpha'}(\vxi')-(\rho,\vxi-\vxi'))}
   \langle P(\vxi',\zeta_{\alpha'})u,v\rangle \langle
   v,P(\vxi,\zeta_\alpha)u\rangle d\vxi d\vxi'.
\end{align}

Summing $\langle A u,v\rangle\langle v,B w\rangle$ over $v$ running
through an orthonormal basis (or averiging over $v$ in the unit
sphere) results in $\langle u, A^\dagger B w\rangle$, so that

\be p^u_T(\bk)=\int\int
\sum_{\zeta_\alpha\in\spectral(\vxi),\zeta_{\alpha'}\in\spectral(\vxi')}e^{iT(\zeta_\alpha(\vxi)-\zeta_{\alpha'}(\vxi')-(\rho,\vxi-\vxi'))}
\langle u, P(\vxi',\zeta_{\alpha'}) P(\vxi,\zeta_\alpha)u\rangle d\vxi
d\vxi'.
\ee

 If the sequence of the measures $\pi^u_T(\rho)$ converges weakly to a
 probability measure (it does, at least along some subsequence, as all
 of these measures are supported on a compact $\ch$), one has
 \begin{align*}
   \lim_{T\to\infty}\sum h(T\rho)\pi_T^u(\rho)&=\lim_{T\to\infty}T^d\int h(\rho) p^u_T(T\rho)d\rho=\\
                                              \lim_{T\to\infty}T^d\int h(\rho)
                                               &\sum_{\zeta_\alpha\in\spectral(\vxi),\zeta_{\alpha'}\in\spectral(\vxi')}
                                                e^{iT(\zeta_\alpha(\vxi)-\zeta_{\alpha'}(\vxi')-(\rho,\vxi-\vxi'))}
                                                \langle u, P(\vxi',\zeta_{\alpha'}) P(\vxi,\zeta_\alpha)u\rangle
                                                d\vxi d\vxi'd\rho=\\
                                              \lim_{T\to\infty}T^d\int h(\rho)
                                              & \sum_{\zeta_\alpha\in\spectral(\vxi),\zeta_{\alpha'}\in\spectral(\vxi')}
                                                e^{iT(\zeta_\alpha(\vxi)-\zeta_{\alpha'}(\vxi+\veta)+(\rho,\veta))}
                                                \langle u, P(\vxi',\zeta_{\alpha'}) P(\vxi,\zeta_\alpha)u\rangle
                                                d\vxi d\veta d\rho
 \end{align*}
     (last line is obtained by the variable change $\vxi'=\vxi+\veta$).

 Now, Duistermaat's trick is to switch the order of the integration,
 performing it first over $\rho$ and $\veta$.  As one
 can easily see, the restrictions of the phases
     \[
       \zeta_\alpha(\vxi)-\zeta_{\alpha'}(\vxi+\veta)+(\rho,\veta)
     \]
     to the $2d$-dimensional spaces of constant $\vxi$ have a unique
     Morse critical point $\veta=0,\rho=d\zeta_{\alpha'}(\vxi)$ of
     index $d$ and and the determinant $1$ for each $\vxi$.  Hence the
     formulae for the asymptotics of the Laplace method apply, localizing the
     integral to the vicinities of those critical points. Using
     further the fact that the projectors $P(\vxi,\zeta), \zeta\in \spectral(\vxi)$ are
     orthogonal at distinct $\zeta\in\spectral(\vxi)$, we derive the limit of the
     integral
       \begin{align*}
         \lim_{T\to\infty}&T^d\int\int\int h(\rho)\sum_{\zeta_\alpha\in\spectral(\vxi),\zeta_{\alpha'}\in\spectral(\vxi')}e^{iT(\zeta_\alpha(\vxi)-\zeta_{\alpha'}(\vxi+\veta)-(\rho,\veta))} \langle u, P(\vxi+\veta,\zeta_{\alpha'}) P(\vxi,\zeta_\alpha)u\rangle  d\veta d\rho d\vxi &=\\
         &(2\pi)^{-d}\int \sum_{\zeta_\alpha\in\spectral(\vxi)} h(d\zeta_\alpha(\vxi)) \langle u, P(\vxi,\zeta_\alpha)u\rangle d\vxi,&
         \end{align*}
   which is equivalent to the claim of the theorem.
\end{proof}

Averaging the probability measures $\pi^u_T(\cdot)$ with respect to
$u$ (again, either over an orthonormal basis, or over the unit sphere
in the space of spins), results in probability measures $\pi_T$
supported on $\ch$.
\begin{cor}\label{cor:ave_u}
  The probability measures $\pi_T$ converge, weakly, to the
  push-forward under the Gauss map of the density on the spectral
  surface equal to $d^{-1} \tr(P(\vxi,\zeta))d\vxi$.
\end{cor}

  \section{Localization}\label{sec:local}
  Localization is a pattern in quantum random walks that attracted significant attention in the literature, see e.g. \cite{inui_localization,ko_one-dimensional_2016,lyu_localization}. Traditionally, strong localization is understood as a nontrivial probability of  return to the initial location: the probability $p^u_T(\zero)$ remains bounded from below. Weak localization just means that the weak limit of $\pi_T^u$ as $T$ goes to infinity has an atom at the origin.

We use a somewhat generalized notion of localization in quantum random
walks:
\begin{dfn}
  The quantum random walk exhibits {\em strong localization} if for
  some initial state $u$, there is a sequence of times
  $T_1,T_2,\ldots$ and states
  $|\bk_1,v_1\rangle,|\bk_2,v_2\rangle,|\bk_3,v_3\rangle,\ldots$ such
  that the sequence of probabilities has nonzero lower limit:
\[
\lim\inf_{k\to\infty} p^u_{T_k}(\bk_k,v_k)>0.
\]
If the sequence of vectors $\bk_k/T_k$ converges to a vector
$s\in\ch$, then we say that there is localization at asymptotic speed
$s$.

The quantum walk localizes {\em weakly} if the limiting measure
$\pi_{u,\coin}$ defined in Theorem \ref{thm:weak} has an atom.
\end{dfn}

In other words, we allow the particle to localize at some point that
moves with linear speed, not necessarily equal to zero. Indeed, nondegenerate affine transformations of the jump vectors commute with taking the weak limits of the respective probability measures, and there is no reason to single out the origin as the localization site.

\subsection{Localizations Strong and Weak}
It is immediate that the strong localization implies that the set of
asymptotic speeds is nonempty (by the compactness of $\ch$ and Tychonov theorem), and that
strong localization implies weak localization.

In \cite{ko_one-dimensional_2016} the equivalence of strong and weak
localizations for a special class of one-dimensional quantum random
walks was proven. In fact, this equivalence is quite general:

\begin{prp}
For translation invariant quantum random walks on lattices, the strong and weak localizations are equivalent.
\end{prp}

To prove this, we use the following corollary of Theorem \ref{thm:weak}.

Define the {\em monomial torus} in $\etorus$ the torus given by
equation $\bz^m\vbx^{\bl}=1$, for some integer $m\neq 0,
\bl\in\Int^d$.

\begin{prp}
The quantum localizes weakly only if the spectral  surface contains a monomial torus as a component.
\end{prp}
\begin{proof}
  Existence of an atom $a\in\ch$ in the limiting measure $\pi$ implies
  that the Gauss map $G$ sends a set of positive measure
  $A\subset\spectral$ to a point. This implies that there is a point
  $(\vxi_*,\zeta_*)\in\smspectral$ on the smooth part of the spectral
  surface, such that $A$ is dense at the point (that is the fraction
  of the volume of a small ball around that point which is in $A$
  tends to $1$ as the radius of the ball tends to zero).

  By Fubini, for almost any $v\in\Real^d,|v|=1$, the curve
  $(\vxi(t),\zeta(t))\in\smspectral, t\in(-\epsilon, \epsilon)$ such
  that $\vxi(t)=\vxi_*+tv$ for small $t$, the intersection of the
  curve with $A$ is dense at $t=0$, and therefore (by analyticity of
  the curve, and algebraicity of the Gauss map), the curve is in $A$
  for all $t$. This implies that in some viscinity of
  $(\vxi,\zeta_*)$, the Gauss map is a constant, and, therefore, again
  by analiticity of $\spectral$, it is constant on an open component
  of $\smspectral$. Thus this component is the level set of a monomial.
\end{proof}

\subsection{Quantization}
This implies
\begin{cor}
If a quantum random walk localizes weakly, the atom belongs to the intersection of sublattice $\frac1k\Int^d, 1\leq k\leq c$ and the jump set convex hull $\ch$.

Also, if the limiting probability measure $\pi_{u,\coin}$ has an atom at $s\in \frac1k\Int^d\cap\ch$, then there is strong localization at the speed $s$.
\end{cor}

In other words, the coordinates of the speeds at which localizations can occur are all
rational, with denominators bounded by the dimension of the chirality
space.

\subsubsection{Example: Standard Hadamard Walk}
As we mentioned, the standard walk with the Hadamard coin, and the jump vectors $(0,0), (0,1), (1,0), 1,1)$ exhibits, numerically, localization patterns at the center of the circle, the image of the spectral surface under the Gauss map (Fig. \ref{fig:hadamard}, left display).

Indeed, the equation of the spectral surface factors:
\[
\det(z\eye-M(\vbx))=1/2 (-2 x y + z + x z + y z + x y z - 2 z^2) (x y - z^2).
\]
The locus of $\{ x y = z^2\}$ consists of two monomial tori, each contributing $1/4$ to the atom at the point $(1/2, 1/2)$.

\section{Random Coin}
Thus far we established (Corollary \ref{cor:ave_u}) that the limiting probability density of the
rescaled position of a QRW corresponding to a coin $\coin$ with the
starting state $|\b0\otimes u\rangle, |u|^2=1$, averaged over $u$, is the image of the
measure $d^{-1} \tr(P(\vxi,\zeta))d\vxi$ on the spectral surface $\spectral$ under the Gauss map.

We will denote this limiting probability measure corresponding to the coin $\coin$ as
$\limmes_{\coin}$.

It is supported, for all coins $\coin$, by the convex hull $\ch$ of
the jump vectors $\bj_i=j(e_i), i=1,\ldots,n$.

It is natural to ask what is the behavior of the
asymptotic measures $\limmes_{\coin}$ for averaged over the coins $\coin$.
Namely,
{\em what is the average of the measures $\limmes_{\coin}$ as $\coin$ is
  distributed over $\su(c)$ according to Haar measure}? While each measure $\pi_T$ is a scintillating pattern, with bring caustics, and even (sometimes) atoms, what is the average behavior of these measures?
  
The answer is surprisingly simple, and is given by the following theorem:

\begin{thm}\label{thm:main}
  The average $\ex_U\limmes_U$ is the pushforward of the uniform
  probability measure on the simplex $\simplex$ spanned by the basis vectors
  $e_i,i=1,\ldots,c$ under the jump map, $\jt:e_i\mapsto \bj(c_i)$, i.e.
$$
\int \limmes_U dU=\jt_*({\mathtt{Uni}}_{\simplex(\basis)}).
$$
\end{thm}

We start the proof with a standard perturbative computation:

\begin{lem}
  If $t\mapsto (\vxi(t),\zeta(t))\in\smspectral$ is a germ of a curve in the
  smooth part of the spectral surface, and the corresponding projector
  $P(\vxi(0),\zeta(0))=|v\rangle\langle v|$ has rank $1$, then
  \be \dot{\zeta}=\langle \Delta(\dot{\vxi}) v,v\rangle, \ee
  (here we
  denote by dot the derivative with respect to the
  parameter on the curve, and by $\Delta(\cdot)$ the diagonal matrix
  with the corresponding vector on the diagonal).
\end{lem}

\begin{proof}
  Under the assumptions, one can choose the eigenvectors smoothly depending on $t$.
  Recall that $\vbx=\exp(i\vxi); \bz=\exp(i\zeta)$.
  Differentiating the identity
  \[
\bz v=\Delta(\vbx)\coin v,
\]
and using the fact that
$|v|^2=1\Rightarrow \langle \dot{v},v\rangle=0$, we obtain
\[
  \dot{\bz}v+\bz \dot{v}=\Delta(\dot{\vbx})\coin
  v+\Delta(\vbx)\coin \dot{v}.
\]
Next we contract this identity with $v$.
Using the equalities
\[
  \Delta(\dot{\vbx})\coin
  v=\Delta(\dot{\vbx})\Delta(\vbx)^{-1}\Delta(\vbx)\coin
  v=i\bz\Delta(\dot{\vxi}) v,
\]
and
\[
  \langle \Delta(\vbx)\coin
  \dot{v},v\rangle=\langle\dot{v},(\Delta(\vbx)\coin)^\dagger
  v\rangle=\langle\dot{v},(\Delta(\vbx)\coin)^{-1} v\rangle=\langle
  \dot{v},\bz^{-1}v\rangle=0,
\]
 we arrive at the desired identity.
\end{proof}

\begin{cor}
The differential of $\zeta$ as a function of $\vxi$ is
given by
\[
d\zeta=\sum_{c} |v_c|^2 (d\vxi,\jt(c))
\]
\end{cor}
\begin{proof}
Direct substitution.
 \end{proof}
 
 In other words, the Gauss map at a smooth point of the spectral surface where the corresponding eigenspace has dimension $1$ is given by the convex combination of the jump vectors, with weights equal to the squared amplitudes of the normalized eigenvector.
 
 The proof of the Theorem \ref{thm:main} follows now from the standard facts:

 \begin{proof}[Proof of Theorem \ref{thm:main}] Consider the average over the coins of the sum of the images of the Gauss map at points of the spectral surface over $\vxi$. For almost all coins, the spectral surface is smooth at those points, and the corresponding eigenspaces one-dimensional. Further, by the unitary invariance of the Haar measure, the distributions of those one-dimensional subspaces will be $\su$-invariant, and therefore the corresponding eigenvectors can be chosen to be uniformly distributed over the unit sphere. As is well-known, the vector of squared absolute values of coordinates (in any orthonormal basis) of a random vector uniformly distributed over the unit sphere is uniformly distributed in the standard simplex.

   This proves that the average (over coins) of the images under the Gauss map of the points of the spectral surface in a given fiber are the push-forward under the jump map of the uniform measure on the standard simplex. As the result is independent of $\vxi$, averaging over $\vxi$ does not change the resulting density.
 \end{proof}
 
\subsection{Simulations}
We conclude with the results of the averaging of the probability density after $T=40$ steps over $1000$ randomly generated unitary coins, for the $d=4$ and the jump maps corresponding to the examples of the Section \ref{subsec:examples}.

One can easily recognize, visually, the resulting densities as the projections of the uniform measure under the corresponding jump maps.

\begin{figure}[h!]
  \begin{center}
\includegraphics[height=1.5in]{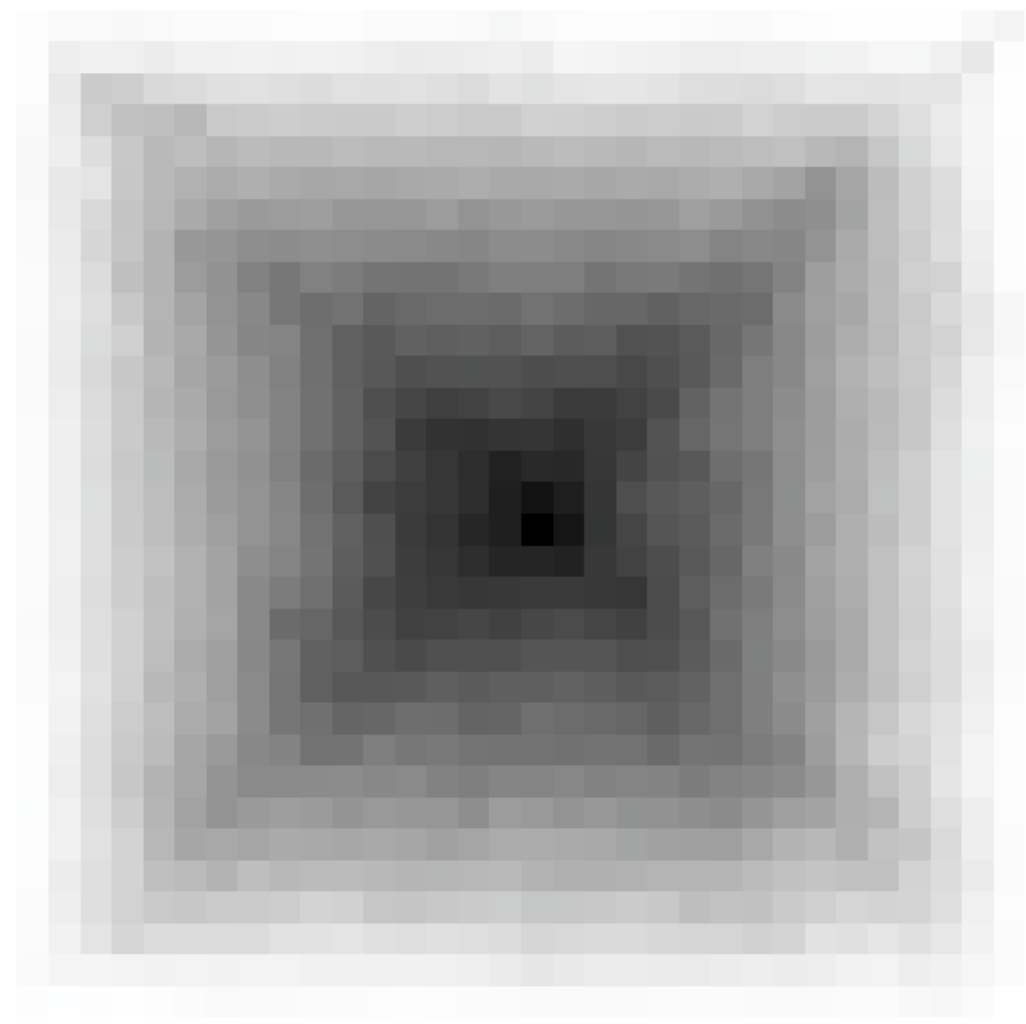}\hfill
\includegraphics[height=1.5in]{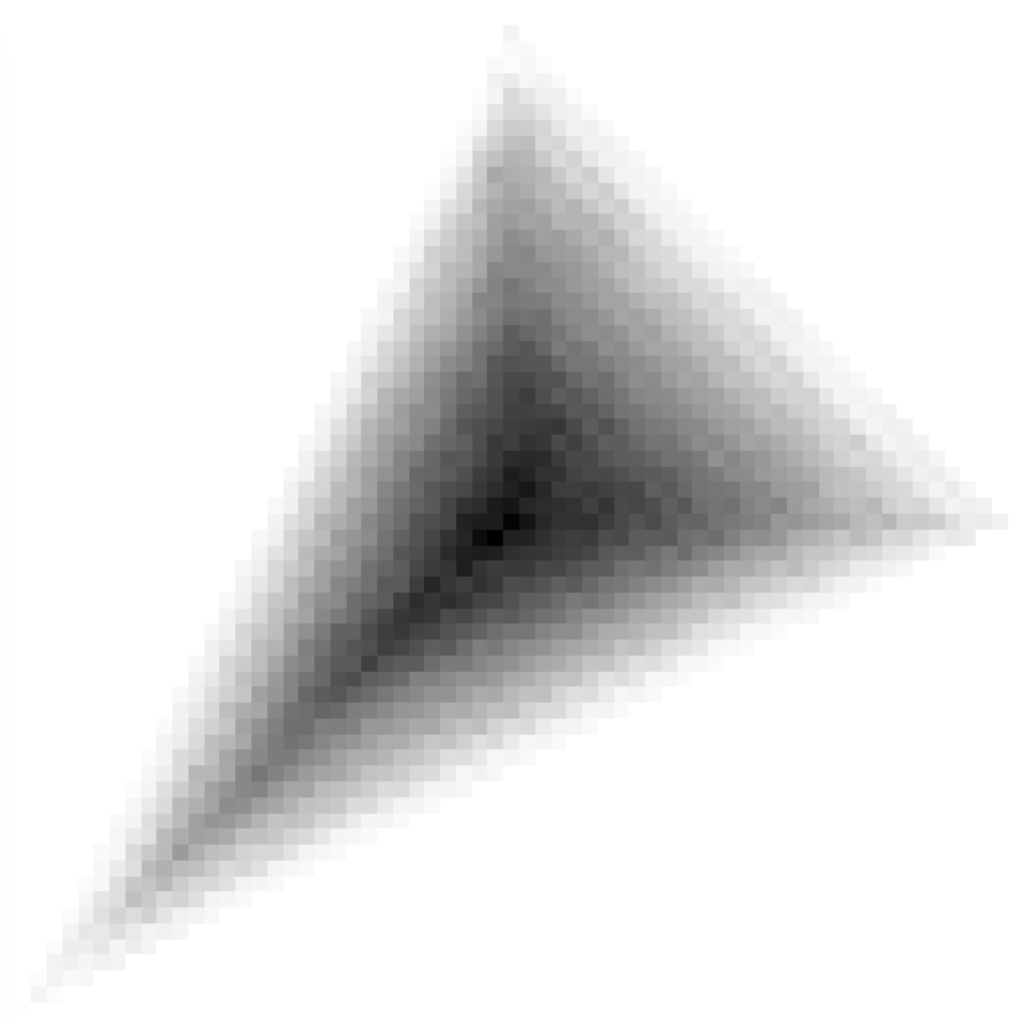}\hfill
\includegraphics[height=1.5in]{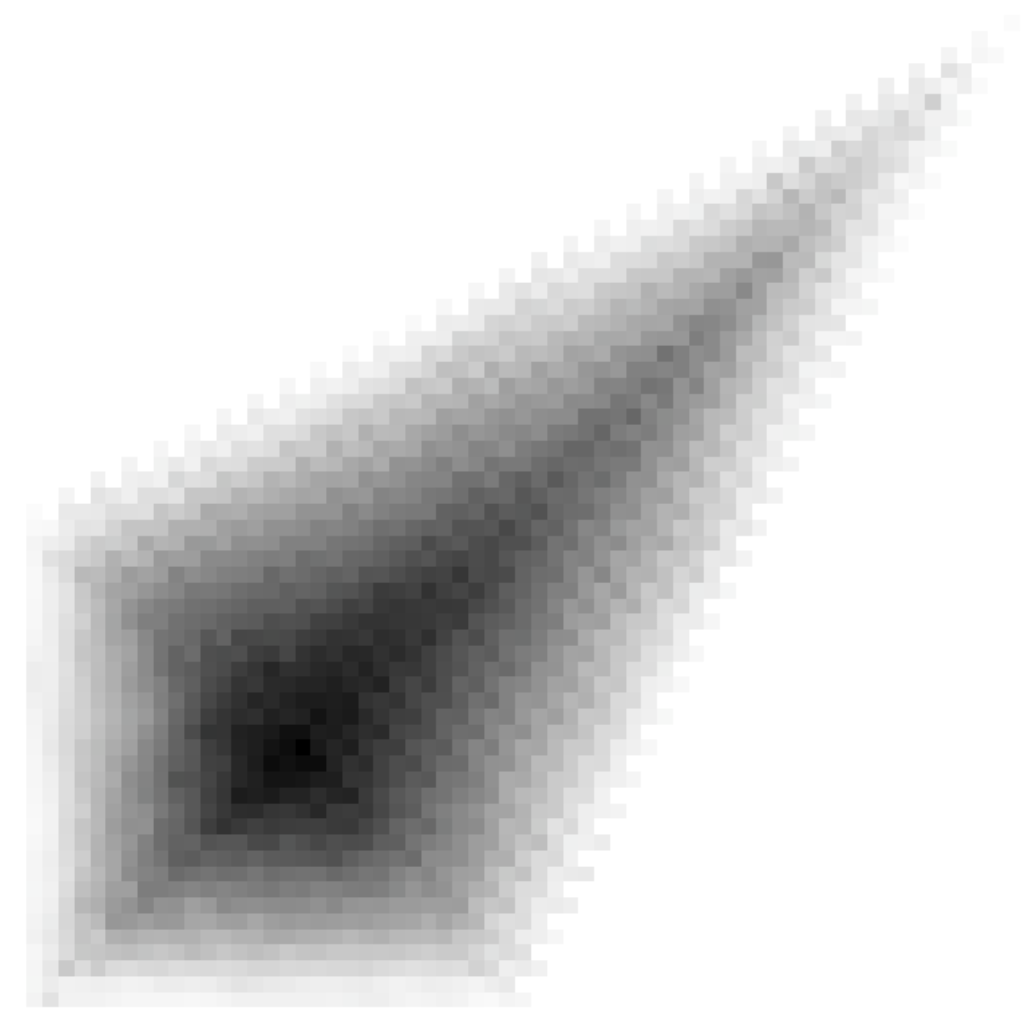}\hfill
\includegraphics[height=1.5in]{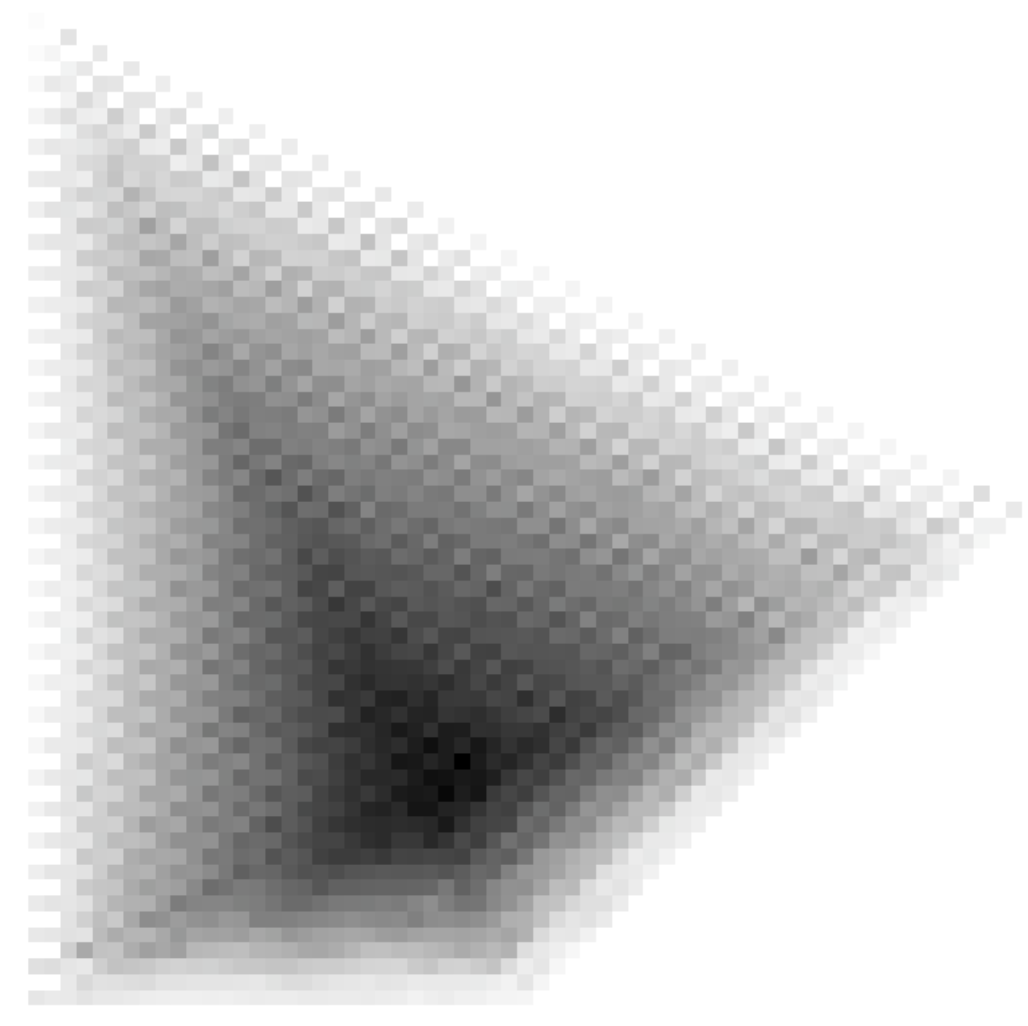}
\caption{Averaged robability distributions for QRWs with jump vectors  $\jt_1,\ldots,\jt_4$, as in section \ref{jump}.
}
\label{fig:average}
\end{center}
\end{figure}

\bibliographystyle{alpha} 
\bibliography{qrw} 

\begin{thebibliography}{AVWW11}

\bibitem[AGZV12]{AVG2}
V.~I. Arnold, S.~M. Gusein-Zade, and A.~N. Varchenko.
\newblock {\em Singularities of differentiable maps. {V}olume 2}.
\newblock Modern Birkh\"{a}user Classics. Birkh\"{a}user/Springer, New York,
  2012.
\newblock Monodromy and asymptotics of integrals, Translated from the Russian
  by Hugh Porteous and revised by the authors and James Montaldi, Reprint of
  the 1988 translation.

\bibitem[AVWW11]{ahlbrecht_asymptotic_2011}
Andre Ahlbrecht, Holger Vogts, Albert~H. Werner, and Reinhard~F. Werner.
\newblock Asymptotic evolution of quantum walks with random coin.
\newblock {\em Journal of Mathematical Physics}, 52(4):042201, April 2011.
\newblock arXiv: 1009.2019.

\bibitem[BBBP11]{baryshnikov_two-dimensional_2011}
Yuliy Baryshnikov, Wil Brady, Andrew Bressler, and Robin Pemantle.
\newblock Two-dimensional {Quantum} {Random} {Walk}.
\newblock {\em Journal of Statistical Physics}, 142(1):78--107, January 2011.

\bibitem[Dui74]{duistermaat}
J.~J. Duistermaat.
\newblock Oscillatory integrals, lagrange immersions and unfolding of
  singularities.
\newblock {\em Communications on Pure and Applied Mathematics}, 27(2):207--281,
  1974.

\bibitem[GJS04]{grimmett_weak_2004}
Geoffrey Grimmett, Svante Janson, and Petra~F. Scudo.
\newblock Weak limits for quantum random walks.
\newblock {\em Physical Review E}, 69(2), February 2004.

\bibitem[IKK04]{inui_localization}
Norio Inui, Yoshinao Konishi, and Norio Konno.
\newblock Localization of two-dimensional quantum walks.
\newblock {\em Physical Review A}, 69(5), May 2004.

\bibitem[KSY16]{ko_one-dimensional_2016}
Chul~Ki Ko, Etsuo Segawa, and Hyun~Jae Yoo.
\newblock One-dimensional three-state quantum walks: {Weak} limits and
  localization.
\newblock {\em Infinite Dimensional Analysis, Quantum Probability and Related
  Topics}, 19(04):1650025, December 2016.

\bibitem[LYW15]{lyu_localization}
Changyuan Lyu, Luyan Yu, and Shengjun Wu.
\newblock Localization in {Quantum} {Walks} on a {Honeycomb} {Network}.
\newblock {\em Physical Review A}, 92(5), November 2015.
\newblock arXiv: 1509.03919.

\bibitem[VA12]{venegas-andraca_quantum_2012}
Salvador~Elías Venegas-Andraca.
\newblock Quantum walks: a comprehensive review.
\newblock {\em Quantum Information Processing}, 11(5):1015--1106, October 2012.

\end{thebibliography}

\end{document}